\documentclass[useAMS,usenatbib]{mn2e}

\usepackage{aas_macros}
\usepackage{graphicx}
\newcommand{\msun}{\mbox{${\rm M}_\odot$}}

\title[The Main Sequence of Star Clusters]{The Main Sequence of Star Clusters}
\author[Andreas H. W. K\"upper, Pavel Kroupa, Holger Baumgardt]{Andreas H. W. K\"upper\thanks{e-mail:
akuepper@astro.uni-bonn.de; pavel@astro.uni-bonn.de; holger@astro.uni-bonn.de}, Pavel Kroupa, Holger Baumgardt\\
Argelander-Institut f\"ur Astronomie, Universit\"at Bonn, Auf dem H\"ugel 71, 53121 Bonn, Germany}

\begin{document}

\date{...}
\pagerange{\pageref{firstpage}--\pageref{lastpage}} \pubyear{2008}
\maketitle

\label{firstpage}

\begin{abstract}
A novel way of looking at the evolution of star clusters is presented. With a \textit{dynamical temperature}, given by the mean kinetic energy of the cluster stars, and a \textit{dynamical luminosity}, which is defined as the kinetic energy of the stars leaving the cluster in analogy to the energy of photons emitted by a star, the dissolution of star clusters is studied using a new \textit{dynamical temperature-luminosity diagram} for star clusters. The investigation contains a parameter-space study of open clusters of up to $N=32768$ single-mass stars with different initial density distributions, half-mass radii, tidal conditions and binary fractions.
The clusters show a strong correlation between \textit{dynamical temperature} and \textit{dynamical luminosity} and most of the investigated cluster families share a common sequence in such a \textit{dynamical temperature-luminosity diagram}. Deviations from this sequence are analyzed and discussed.
After core collapse, the position of a cluster within this diagram can be defined by three parameters: the mass, the tidal conditions and the binary fraction. Due to core collapse all initial conditions are lost and the remaining stars adjust to the given tidal conditions. Binaries as internal energy sources influence this adjustment.
A further finding concerns the Lagrange radii of star clusters: Throughout the investigated parameter space nearly all clusters show a constant half-mass radius for the time after core collapse until dissolution. Furthermore, the ratio of half-mass radius to tidal radius evolves onto a common sequence which only depends on the mass left in the cluster.
\end{abstract}


\section{Introduction}\label{sec:intro}
In 1913 Henry Norris Russell presented his work on a relation between the spectral classes of stars and their absolute magnitude at a meeting of the Royal Astronomical Society \citep{Ru13}. The diagram he showed later became famous as the Hertzsprung-Russel Diagram (Ejnar Hertzsprung was the first to anticipate the existence of a relation between the two quantities) and became one of the most important tools for the study and understanding of stellar evolution.

The temperature-luminosity diagram, which is a derivative of the original Hertzsprung-Russel diagram, shows a tight relation between a star's temperature and its luminosity for the main-sequence phase of stars.

A star cluster is a system which shows some analogies to a star. First of all, the stars within a cluster follow a velocity distribution, which is established through two-body encounters, just like a gas or plasma does through collisions of particles. Therefore it is possible to assign a \textit{dynamical temperature}, $T$, to the stars in a cluster (Sec.~\ref{sec:T}).
Secondly, a star cluster constantly loses a certain fraction of its stars through escape, like a star constantly emits photons. It thus appears plausible to define a \textit{dynamical luminosity}, $L$, in terms of the energy carried away by the stars per unit time (Sec.~\ref{sec:L}).

To find if a dynamical temperature-luminosity relation exists for dynamical systems like star clusters is the motivation of this work, because it is likely that it would prove very useful for describing global cluster properties and evolution. Using numerical simulations, the behaviour of star clusters within a \textit{dynamical T-L diagram} (Sec.~\ref{sec:TL}) and the influence of different initial conditions on its development (Sec.~\ref{sec:var}) is studied.

Throughout the performed parameter-space study the half-mass and tidal radius of the clusters are investigated in detail, as they turned out to not behave as expected. The half-mass radius stays constant for a large fraction of a cluster's life-time and the ratio of half-mass radius to tidal radius evolves along a single sequence, independent of initial conditions.

But first of all a reference model, the ``\textit{Standard Cluster}'', is defined and investigated in detail (Sec.~\ref{sec:scl}), which will help establishing a \textit{dynamical temperature} and a \textit{dynamical luminosity}.

\section{Standard Cluster}\label{sec:scl}
Since the number of possible initial models for N-body computations of star clusters is infinite, it is necessary to scan the available parameter space systematically. Therefore it is highly convenient to define a ``\textit{Standard Cluster}'' with carefully chosen, plausible initial conditions, which is taken as a reference model. After detailed studies of this type of model, the effects of changes to the initial conditions can be traced efficiently. From these studies of the different variations, general features of cluster evolution can be deduced.

For this investigation about one hundred and fifty clusters with different initial conditions were calculated. They all were computed with the $N$-body code NBODY4 \citep{Aa99, Aa03} until total dissolution (i.e. until $N=10$). Although NBODY4 is capable of including stellar evolution and similar effects, they were not implemented here to focus on pure dynamical evolution. 

It turned out that for clusters with up to $\sim$2000 stars, a regular workstation is sufficiently fast. The larger models were computed on the GRAPE-6A special purpose computers of the Argelander Institute \citep*{Fu05}.\\

The \textit{Standard Cluster} is defined as follows:
\begin{enumerate}
 \item Plummer radial density profile,\label{pp}
 \item half-mass radius of 0.8 pc,\label{rh}
 \item 1000 equal-mass stars of 1 $\msun$,\label{nr}
 \item tidal field,\label{tf}
 \item no primordial binaries.\label{bn}
\end{enumerate}
Details to the several choices are given below. 

\begin{enumerate}
\item The Plummer profile is chosen because of its analytical convenience \citep*{Aa74, He03}.

\item The choice of $R_h=0.8\,\mbox{pc}$ is motivated by \cite{Kr95}, who finds by inverse dynamical population synthesis that the majority of Galactic field stars is probably born in aggregates of this size. 

\item $N=1000$ is a good compromise between statistics and computing time; on a regular workstation such a cluster can be computed within a few hours.

Constraining the masses of the stars to one solar mass puts the focus on the pure dynamical evolution of the clusters, since stellar evolution needs not to be treated. Furthermore it avoids further parameters which have to be fixed initially (e.g. the initial mass function). 

\item The tidal field is added in the near field approximation, which is a linearisation of the galaxy potential \citep*{Ro97}. Here it is assumed that the cluster moves on a circular orbit in the Galaxy, which is treated as a point mass $M_G = 9.565 \cdot 10^{10}\,\msun$ at distance $R_G = 8.5\,\mbox{kpc}$, corresponding to a rotational speed of 
$v_{rot}=220 \mbox{km}\mbox{s}^{-1}$. The \textit{Standard Cluster} is therefore subject to boundary conditions typical for clusters in the solar neighbourhood. 

The computations are conducted in a rotating reference frame, which moves with the cluster's initial angular velocity, with the origin of the coordinate system lying in the initial barycentre of the cluster. The $x$-axis points to the Galactic centre and the $y$-axis is parallel to $\vec{v}_{rot}$, while the $z$-axis is perpendicular to the orbit of the cluster. For further details see \citet{Fukushige00}. 

All models start in dynamical equilibrium but the tidal field is added after setting up the initial positions and velocities of the stars. Thus a few stars may already be unbound at the beginning. Because the initial clusters are quite small compared to the tidal radius this effect may be neglected, but becomes more important for initially very extended clusters. 

\item Binaries slow down computations enormously because time steps have to be very small for their integration. Hence, although \citet{Kr95} suggests a rather large primordial binary fraction, the models did not contain primordial binaries, i.e. the primordial binary fraction, $f_{bin}$, given by
\begin{equation}\label{eq:fbin}
 f_{bin} = \frac{N_{bin}}{N_{bin}+N_s},
\end{equation}
where $N_{bin}$ is the number of binary systems and $N_{s}$ is the number of single stars, such that $N=2N_{bin}+N_s$, was set to zero. 
\end{enumerate}

Since the \textit{Standard Cluster} offers little statistics due to its small $N$, accurate results on statistical properties can only be found by combining many renditions of the same model, differing only in the initial positions and velocities of the stars. Therefore in the following sections the insights from simulations of 27 renditions of the \textit{Standard Cluster} are presented, eventually supported by exemplary outcomes from other models. In Sec.~\ref{sec:var} each variation is discussed in detail.

\section{Dynamical Temperature}\label{sec:T}
\begin{figure}
  \includegraphics[width=84mm]{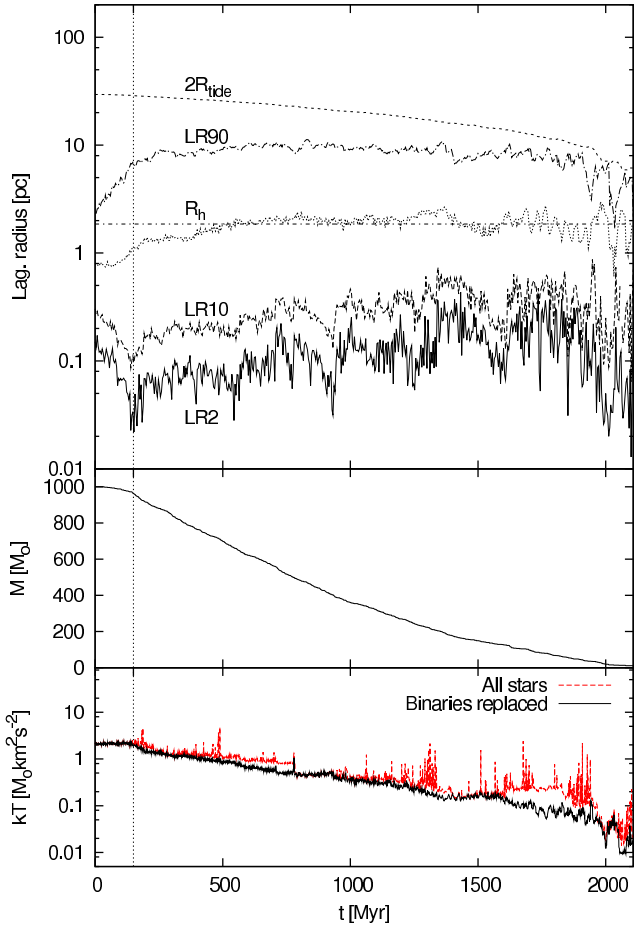}
  \caption{Upper panel: Lagrange radii of a \textit{Standard Cluster}; LR2, LR10 and LR90 are respectively the 2\%, 10\% and 90\% Lagrange radii (the vertical line shows $t_{cc}$, the horizontal line is the fitted mean value of the half-mass radius, $R_{h}^f$, for $t > t_{cc}$). While the core undergoes oscillations, the outer shells expand until they reach a nearly constant state (especially the half-mass radius). The tidal radius (dotted curve) decreases continuously due to ongoing mass loss. The middle panel shows the mass evolution, and the bottom panel the temperature evolution with and without replacing the binaries with their centre-of-mass particles. After rising slightly, the temperature drops after core collapse. Without replacement, binary systems dominate the cluster temperature.}
  \label{dMdiffsc}
\end{figure}

\begin{figure}
\includegraphics[width=84mm]{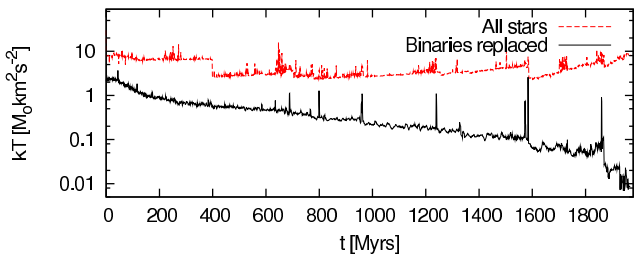}
  \caption{Temperature evolution of a cluster with the same properties like a \textit{Standard Cluster}, except $f_{bin}=0.95$, plotted with and without correcting for binaries. Tight binaries completely dominate the cluster temperature when they are not replaced by the corresponding centre-of-mass particles.}
  \label{Tbin}
\end{figure}

A powerful analogon to the dynamics of a cluster with stars of a single mass is given by the kinetic theory of gases \citep{Louis91}. In this framework the cluster is treated as an ideal gas with point-particles interacting only through collisions. Although this analogy exhibits some flaws, since interactions between stars are not confined to the short moment of collision, one can get valuable quantitative results out of it.

In analogy to an ideal gas, a \textit{dynamical temperature} of a cluster can be defined through the mean velocity of the member stars. The three-dimensional squared velocity dispersion is given by 
\begin{equation}\label{eq:sigma}
 \sigma^2=\frac{1}{N}\sum\limits_{i}\left(\left(v_{xi}-\overline{v}_x\right)^2+\left(v_{yi}-\overline{v}_y\right)^2+\left(v_{zi}-\overline{v}_z\right)^2\right),
\end{equation}
where $N$ is the number of bound stars (which will be defined in Sec.~\ref{sec:L}), the $v$'s are the velocity components and the $\overline{v}$'s the corresponding mean velocities. For a star cluster at rest,
\begin{equation}
 \overline{E}_{kin} = \frac{1}{2}m\overline{v^2} = \frac{1}{2}m\sigma^2,
\end{equation}
where $m$ is the mass of the stars, which is taken to be the same for all stars, and $\overline{E}_{kin}$ is their mean kinetic energy. Furthermore, it is possible to define the pressure of a star cluster,
\begin{equation}\label{eq:P}
 P = \frac{Nm\sigma^2}{2V},
\end{equation}
where $N$ is the number of particles and $V$ the volume of the system. With the so found quantity $P$ and the ideal gas equation,
\begin{equation}
 PV = NkT,
\end{equation}
where $k$ is Boltzmann's constant, a temperature $T$ can be defined, which is proportional to the squared velocity dispersion of the cluster, i.e. the mean kinetic energy of the member stars:
\begin{equation}
 \frac{1}{2}m\sigma^2 = kT.
\end{equation}
In the case of a star cluster the temperature is usually denoted as $kT$, thus has the dimension of an energy. Like for a perfect gas, the temperature therefore gives a direct idea of the distribution of velocities within the cluster.

But this approximation breaks down when two particles form a binary, since then the premise of an ideal gas is not valid any more. A single tight binary is able to completely dominate the cluster temperature, due to the large internal stellar motion, and distort the frame of kinetic theory. Therefore these systems have to be neglected for the measurement of the temperature. 

Taking a look at the actual binary content of a \textit{Standard Cluster} reveals a rather small fraction of cluster stars captured in binaries, i.e. the number of binaries in a \textit{Standard Cluster} is mostly less than ten which validates the above approximations (see also Fig.~\ref{vt} in Sec.~\ref{sec:L}). Hence, for the temperature analysis, i.e. in eq. \ref{eq:sigma}, binaries are replaced by their centre-of-mass particles.

This corrected temperature is a direct measure of the velocity distribution of the stars within the cluster. This distribution is supposed to be Maxwellian, which implies that there is a high-velocity tail of stars, and, since a star cluster has a finite escape speed, that a number of stars are unbound. When these hot stars have left the cluster, the left-over stars will reestablish a new velocity distribution with a lower temperature within a two-body relaxation time. This process continues until the cluster has dissolved. This can be seen in the lower panel of Fig.~\ref{dMdiffsc}, where the value of $kT$ of a \textit{Standard Cluster} in the course of time is shown with and without replacing the binaries. 
In this context the importance of the temperature correction becomes clear when looking at the temperature evolution of a cluster with a substantial fraction of primordial binaries (Fig.~\ref{Tbin}). Without replacing the binaries by their centre-of-mass particles, kinetic theory completely breaks down, since the measured temperature is not related to the actual velocity distribution of the stars within the cluster. In this particular example, the temperature of the cluster at dissolution time would be even higher than in the beginning, which is to be expected due to the negative heat capacity of gravity.

The temperature evolution of the \textit{Standard Cluster} can be explained by looking at the top and middle panel of Fig.~\ref{dMdiffsc}. In the beginning the cluster is very compact and sits deeply within the tidal sphere. As the core immediately starts collapsing (top panel) the temperature rises and the outer layers expand. Core collapse is reached at the indicated time, $t_{cc}$, where the \textit{Standard Cluster} has expanded sufficiently to fill its tidal radius and starts spilling over the tidal radius. This can also be observed by the change of slope of the mass evolution (middle panel).

A further feature in the top panel of Fig.~\ref{dMdiffsc} is striking: after core collapse the half-mass radius is nearly constant for the rest of the cluster's life-time. This feature can be observed in almost all computed clusters and will be discussed in detail in Sec.~\ref{sec:var}. The clusters adopt to the tidal conditions and settle to an equilibrium half-mass radius, where energy production in the core (which causes cluster expansion) and energy loss at the tidal radius balance each other. The dashed horizontal line in  Fig.~\ref{dMdiffsc} gives a fit to the half-mass radius after core collapse until dissolution time, $t_{dis}$, of this particular cluster. This has been done for all computed models. As can be seen in Tab.~\ref{table1} (column $R_h^f$), all clusters with $N = 1000$ and $R_{Gal} = 8.5$ kpc establish a half-mass radius of about 2 pc.

\section{Dynamical Luminosity}\label{sec:L}
\begin{figure*}
  \includegraphics[width=84mm]{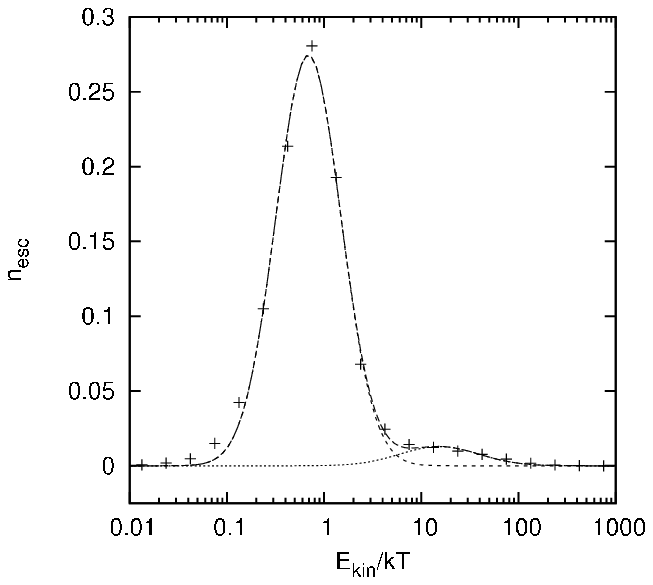}\hfill
  \includegraphics[width=84mm]{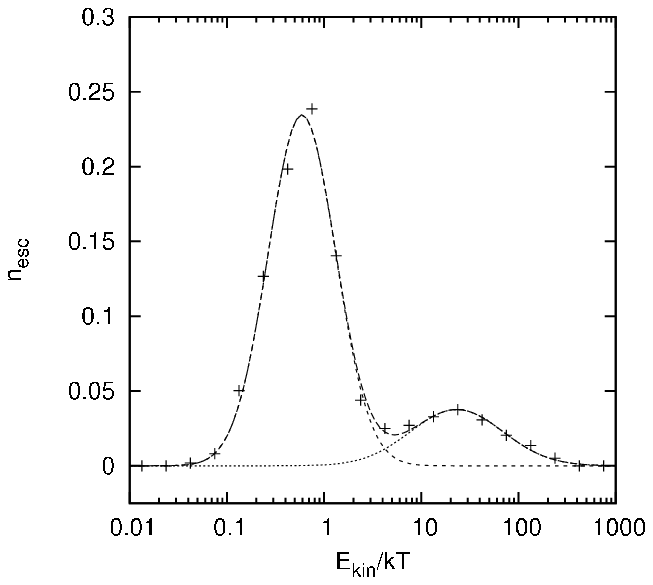}
  \caption{Distribution of mean normalised escape velocities of 27 \textit{Standard Clusters} (left) and 9 isolated clusters (right). $n_{esc}$ gives the probability of detecting an escaper with the given $E_{kin}/kT$. The dotted lines show the fitted lognormal distributions which intersect at about $E_{kin}/kT=5.1$. In isolated clusters evaporation is less common than in \textit{Standard Clusters}.}
 \label{1000fit}
\end{figure*}

\begin{figure}
  \includegraphics[width=84mm]{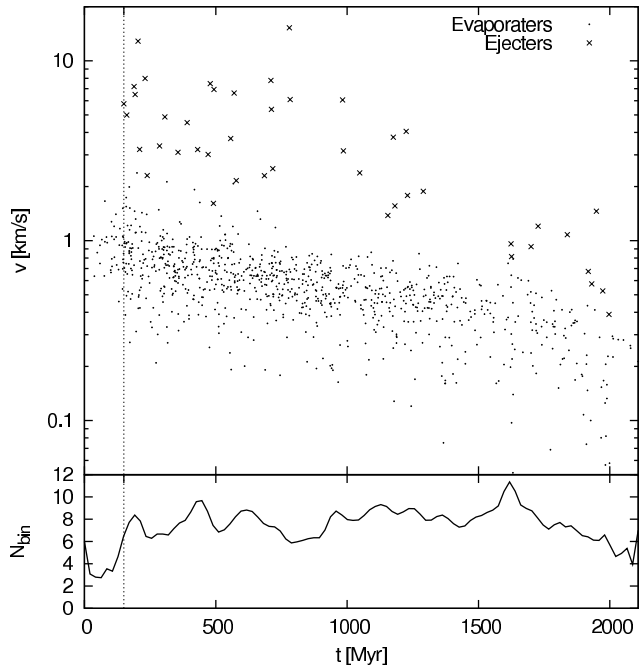}
  \caption{Upper panel: velocities of escaping stars as a function of time. While the majority of escapers leave the cluster with a velocity comparable to the escape velocity and $E_{kin}/kT < 5.1$ (``\textit{Evaporaters}''), a few stars are virtually shot out of the cluster (``\textit{Ejecters}''). The dotted vertical line gives the time of core collapse, $t_{cc}$, which coincides with the occurrence of the first \textit{ejecter}. Lower panel: smoothed number of bound binary systems. During core collapse, binaries are formed which then start producing \textit{ejecters}.}
  \label{vt}
\end{figure}

In general the luminosity of a system can be defined as the amount of energy emitted in a given time interval. For a star this energy is given by the number of emitted photons and their energy. 

When trying to define a \textit{dynamical luminosity} for star clusters in terms of kinetic energy carried away by escaping stars, it is first necessary to define when a star has left the cluster. Answering this question is not trivial since stars can still be bound even beyond the tidal radius or when they already have the necessary energy for escape \citep{Ba01}. A sufficient and simple escape criterion is \citep{Aa03}
\begin{equation}
  \vert\vec{x}_i-\vec{x}_d\vert > 2 R_{tide},
\end{equation}
where $\vec{x}_i$ is the star's position, $\vec{x}_d$ is the cluster's density centre and $R_{tide}$ is the tidal radius. The latter can be calculated as \citep{Spitzer87}
\begin{equation}\label{eq:rtide}
 R_{tide}=R_G\left( \frac{M}{3M_G}\right)^{1/3}.
\end{equation}

Hence, the easiest way of measuring a cluster's luminosity is to put a virtual sphere around the cluster with radius 2 $R_{tide}$ and record the kinetic energy of each star leaving this sphere within a given time interval. In this way the luminosity is determined on the ``surface'' of the cluster. This also implies that all quantities, like the temperature, are most appropriately determined for all stars within this sphere.

Investigating the velocities of the escaping stars reveals two distinct families of escapers \citep*{Gi94b,Ba02}. In the left panel of Fig.~\ref{1000fit} the distribution of the kinetic energies of the escaping stars of all computed \textit{Standard Clusters} is shown. Here the energies have been normalised by the current temperature $kT$ of the cluster, which is reasonable to do when trying to compare different kinds and states of clusters, because in a hotter system the mean velocity is shifted to higher values and so is the mean escape velocity. In this plot $n_{esc}$ is normalised such that 
\begin{equation}
\sum\limits_i n_{esc,i}=1.
\end{equation}
Two lognormal distributions which overlap at about $E_{kin}/kT=5.1$ are readily apparent. The origin of these two families of escapers lie in the way they are produced. 

\begin{enumerate}
\item The slow escape is usually called evaporation. It is due to a number of weak encounters, which gradually increase the star's energy until it reaches a velocity higher than the escape velocity. Evaporation therefore is a consequence of the high-velocity tail of the velocity distribution of stars within a cluster, which is continuously established through two-body relaxation. Since, for a given cluster, the relaxation time decreases with increasing temperature and vice versa, it is supposed that the evaporation rate is correlated to the temperature.

\item High-velocity escapers are due to ejection processes. These can be encounters between a star and a binary or two binaries, where a binary system is hardened while the excess energy is transferred to the other star(s). This implies that the energy taken away by \textit{ejecters} is mostly stored in a remaining binary system, unless this binary recoils sufficiently strongly to leave the cluster. As the stars have to come very close to each other, these events are quite rare and mostly happen in the cluster core.
\end{enumerate}

For further considerations it is convenient to split escapers into \textit{ejecters} and \textit{evaporaters}. For the \textit{Standard Cluster} a threshold $\left(E_{kin}/kT\right)_{limit}$ can be determined from the intersection point of the two lognormal distributions. 
Of course, this also has to be checked for clusters with different intital conditions. For all computed classes of models normalised velocity distributions were acquired, all resembling the distribution of the \textit{Standard Cluster}: Here also two lognormal distributions were fitted and their intersection point was determined. The individual results can be found in Tab.~\ref{table1}. 

Since all models show a limit of about the same value like the \textit{Standard Cluster}, a weighted mean of these values is suggested as the threshold between \textit{evaporaters} and \textit{ejecters},
\begin{equation}
 \left(\frac{E_{kin}}{kT}\right)_{limit} = 5.1 \pm 0.2.
\end{equation}
The same analysis has even been done for the computed isolated clusters, where no tide removes the slow stars, hence evaporation is less efficient (see \citealt{Ba02}). In Fig.~\ref{1000fit} the relative increase of \textit{ejecters}, in number and also in relative energy, can be seen. The fit yields a limit of $E_{kin}/kT = 4.3 \pm 0.6$ which agrees within the errorbars with the results for clusters in tidal fields and therefore seems to be independent of the tidal conditions. 

Looking at the velocities of escaping stars reveals that most ejections take place at and right after core collapse (see Fig.~\ref{vt}), which is highly correlated to the formation of binaries. The probability of a close encounter is highest in this phase. 

In fact, the first \textit{ejecter} of a star cluster can be taken as a trigger for core collapse, i.e. just at the moment when the first escaper with $E_{kin}/kT > 5.1$ is detected, the core has reached its densest phase. This criterion should be preferred, because it is less arbitrary than the usual way of defining $t_{cc}$, which is the formation of the first hard binary with a given amount of binding energy \citep{Gi94b}. The first \textit{ejecter} on the other hand marks a definite change in the core energy evolution, since a large amount of energy is removed immediately. 

The considerations made so far suggest the definition of a dynamical total luminosity, $L_{tot}$, which is given by the kinetic energy carried away by escaping stars. Since two distinct processes are at work it is reasonable to split this total luminosity into an \textit{evaporation luminosity}, $L_{ev}$, and an \textit{ejection luminosity}, $L_{ej}$, such that
\begin{equation}
 L_{tot} = L_{ev} + L_{ej}.
\end{equation}
The \textit{evaporation luminosity} of a cluster is comparable to the luminosity of a star, where photons carry away heat with energies proportional to their frequency. So, to stay in the framework of stars, \textit{evaporaters} may be the analogon to black-body radiation, while \textit{ejecters} are more like solar flares. Evaporation takes place constantly and the amount of energy being lost by it strongly depends on the cluster's temperature, much like the luminosity of a main-sequence star correlates with its effective temperature. Ejection is mostly due to structural effects happening deep inside the cluster, such as core collapse, thus due to the formation and hardening of binaries. It therefore happens more randomly, but is a good tracer for binary activity. 

\begin{figure}
  \includegraphics[width=84mm]{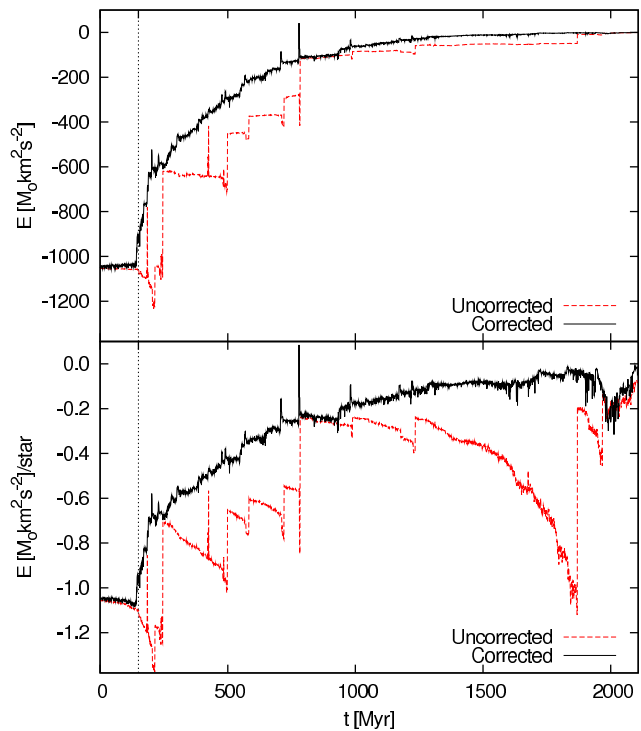}
  \caption{Upper panel: Total energy of the cluster with (thin dashed curve) and without (thick solid curve) counting the internal energy of binaries (binaries being replaced by centre-of-mass particles). Lower panel: total energy per star. From being tightly bound in the beginning, the cluster dissolves completely. While stars constantly leave the cluster the binding energy is predominantly stored in binary systems. The jumps in energy coincide with binary escapers which take away large fractions of the total binding energy. The dashed vertical line marks core collapse.}
  \label{E}
\end{figure}

But a last paradoxon still has to be solved: if an escaping star carries away positive energy and the total energy is negative in the beginning, the cluster should get increasingly bound in the course of time. How come that clusters dissolve finally? The answer is that the binding energy is redistributed among the left-over stars predominantly such that binaries are formed or hardened. These binaries store binding energy until they are removed from the cluster due to a three- or more-body event in the cluster core. 
In Fig.~\ref{E} the energy evolution of a \textit{Standard Cluster} can be seen. The jumps indicate binary-escaper events, which all coincide with a high-velocity \textit{ejecter} of more than 10 km/s going in the opposite direction of the binary-escaper. This supports the argument that \textit{ejecters} are strongly correlated to binary activity. As can be seen in the figure, by simply counting the steps in the energy evolution, this particular \textit{Standard Cluster} produces 12 binary escapers within its life-time (compared to about 40 \textit{ejecters} and more than 900 \textit{evaporaters}). Hence binary escapers are too rare to define a separate luminosity. So, if a binary escapes, just the kinetic energy of its centre of mass is measured and added to the appropriate luminosity.

\section{The Temperature-Luminosity Diagram}\label{sec:TL}
\begin{figure}
    \includegraphics[width=84mm]{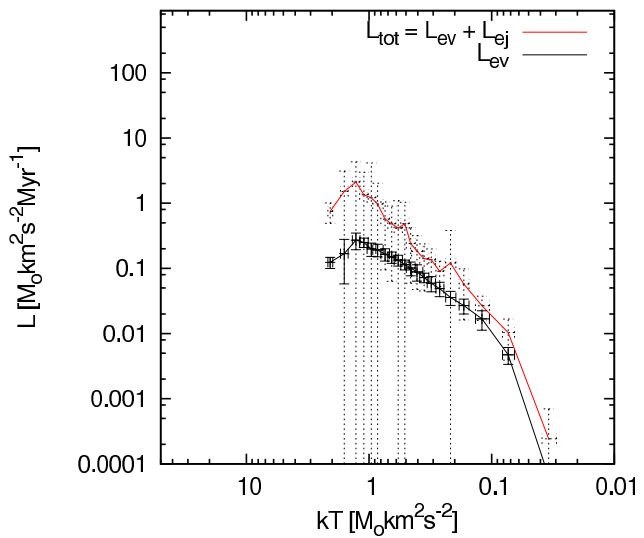}
  \caption{Mean temperature-luminosity diagram of all 27 \textit{Standard Clusters} taking only evaporation into account ($L_{ev}$) and taking both, evaporation and ejection, into account ($L_{tot}$). Errorbars represent the standard deviation from the mean. Since ejection only takes place in dense environments, it is dominant in the beginning but decreases as the cluster dissolves.}
  \label{1000tl}
\end{figure}

Having now introduced the \textit{dynamical temperature}, $kT$, and the \textit{dynamical luminosity}, $L$, a Hertzsprung-Russel-type diagram for star clusters can be drawn (Fig.~\ref{1000tl}). 

Taking only escapers with $E_{kin}/kT < 5.1$ into account and averaging over all 27 \textit{Standard Clusters} yields a smooth curve in the \textit{dynamical temperature-luminosity diagram}. The \textit{Standard Cluster} starts deep within its tidal radius with an \textit{evaporation luminosity} of about 0.1 $\msun\mbox{km}^2\mbox{s}^{-2}\mbox{Myr}^{-1}$ and a temperature of $\sim 2 \msun \mbox{km}^2\mbox{s}^{-2}$, then increases its luminosity before core collapse, after which it cools down constantly until it has dissolved completely to a temperature of about zero. 

The total luminosity, including evaporation and ejection, is a factor of 10 higher and is governed by single \textit{ejecters}. It therefore does not yield a smooth curve but shows a variety of peaks which are of statistical nature. For more massive clusters, which show pronounced core oscillations, these peaks should be correlated to these oscillation, because during contraction phases the central density, and hence the probability of few-body encounters, is highest.

Since in $N$-body models with single-mass particles the primary parameter for defining the state of the cluster is the total mass rather than the age, the different renditions are binned, combined and averaged at equal masses in steps of 50 $\msun$. The errorbars give the standard deviation of single models from the mean. As is evident, the \textit{ejection luminosity} suffers from much stronger statistical fluctuations than the \textit{evaporation luminosity}.

Finally, which deductions can be drawn from the computations of the \textit{Standard Cluster}? First of all, a reasonable way of defining a \textit{dynamical temperature} and a \textit{dynamical luminosity} have been found. This allows to set up a \textit{dynamical temperature-luminosity diagram} and look at cluster evolution in a completely new way. The individual \textit{Standard Clusters} evolve along a single sequence within this diagram, which is in this case more comparable to a cooling sequence of white dwarfs than to a main sequence. This is due to the cluster temperature, which is constantly declining, except for the period before core collapse.

Furthermore, as can be seen in the different slopes of the two curves in Fig.~\ref{1000tl}, energy loss through ejection is most important around core collapse and then decreases compared to evaporation. This difference is due to a different dependence of the two luminosities on cluster parameters such as the number of stars and the density. While the \textit{ejection luminosity} is mainly related to the central velocity dispersion and density, because binaries are most active in the core, the \textit{evaporation luminosity} depends on the total density distribution, because \textit{evaporaters} are generated everywhere in the cluster \citep{Ba02}.

After this pioneering work, the variations of the initial parameters and their influences on the cluster evolution within the \textit{dynamical temperature-luminosity diagram} can be traced.

\section{Variations}\label{sec:var}
\begin{figure}
  \includegraphics[width=84mm]{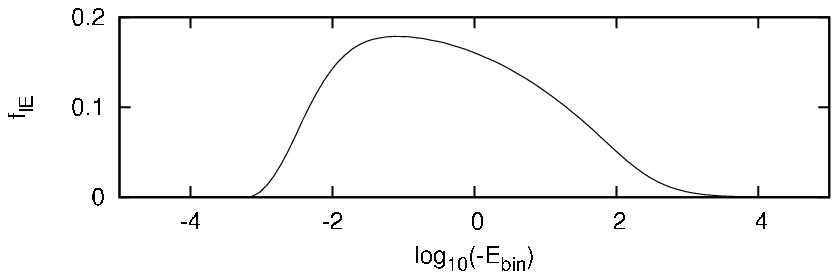}
  \caption{Binding energy distribution function $f_{lE}$ (eq. \ref{flE}). The binding energy is given in $\msun\mbox{km}^2\mbox{s}^{-2}$.}
  \label{flEvergl}
\end{figure}

The following variations of the \textit{Standard Cluster} were investigated (the full list of model properties is given in Tab.~\ref{table1}): 

\begin{enumerate}
\item As an alternative to the Plummer profile, several models with an initial King profile ($W_0 = 5$) were computed \citep{Ki66}. These models were chosen to have the same initial half-mass radius like the \textit{Standard Cluster} of 0.8 pc, hence they are tidally underfilling by a factor of three.

\item The half-mass radius was varied to 0.4, 1.6, 2.4, 3.2 and 4.0 pc.

\item The number of stars was varied from 500 up to 32768 stars, corresponding to 500-32768 $\msun$. The effect of adding a mass function has to be studied in some future work, but is expected to have a large influence on cluster evolution (see \citealt{He69,vanAlbada68,delaFuenteMarcos95}).

\item Models were calculated at a larger Galactic radius of $R_G=85$ kpc. For these models the tidal radius is about 60 pc, which is five times larger than for a \textit{Standard Cluster}. The cluster therefore can expand much further which, at the same time, increases its dynamical time scale significantly.

Furthermore isolated models were computed to get a better idea of the general influence of a tidal field.

\item To see the effect of a substantial binary fraction, $N$-body integrations with $f_{bin}=0.95$ were performed. No other binary fractions were investigated to keep the number of models containing primordial binaries to a minimum because of the long computation times for such models, which are up to 100 times larger than for models without primordial binaries.

To minimise the changes from model to model with regard to future experiments, where a realistic multi-mass spectrum will be implemented, the binding energy distribution, $f_{lE}$, was specified in such a way that the binding energies of the binary systems follow the observed distribution of real binaries, but here applied to single-mass systems.

For a realistic binary system, take two stars $m_1$ and $m_2$ independently out of an initial canonical mass distribution $\xi(m)\propto  m^{-\alpha_i}$ \citep{Kr01} with
\begin{eqnarray}
    \alpha_1 &=& 1.3 \quad\mbox{for}\quad 0.08 \leq m/\msun < 0.5,\label{alpha1} \\
    \alpha_2 &=& 2.3 \quad\mbox{for}\quad 0.5 \leq m/\msun < 150.\label{alpha2}
\end{eqnarray}
Chose a binary period $P$ which follows the period distribution $f_{lP}$ derived by \citet{Kr95a},
\begin{equation}
       f_{lP} = 2.5 \frac{lP-1}{45+(lP-1)^2} \quad\mbox{with}\quad 1 \leq lP \leq 8.43,\label{flP}
\end{equation}
where $lP\equiv \log_{10}(P)$ and $P$ is given in days in this equation. The semi-major axis, $a$, becomes
\begin{equation}
       \frac{a^3}{P^2} =  \frac{G(m_{1}+m_{2})}{4\pi^2}.
\end{equation}
Finally, the binding energy of the binary is
\begin{equation}
       E_{bin}  =  -\frac{G m_1 m_2}{2 a}.
\end{equation}
The distribution of binding energies is then given by a triple integral over the mass ranges of the two composite stars and over the above given range of periods,
\begin{equation}\label{flE}
 f_{lE} = \int\limits_{m_1}\int\limits_{m_2}\int\limits_{lP}f_{lP}\,\xi(m_1) \xi(m_2) \delta(lP-\widetilde{lP})\,dm_1 \,dm_2 \,dlP,
\end{equation}
with the substitution
\begin{equation}
    \widetilde{lP} =  \log\left(\frac{\pi G}{\sqrt{2}}\right)+\frac{1}{2}\log\left(\frac{(m_1 m_2)^{3}}{m_1+m_2}\right)-\frac{3}{2}\,lE,
\end{equation}
where $lE\equiv \log_{10}(\vert E_{bin}\vert)$. The result is a distribution function for binding energies, $f_{lE}$, combining an initial mass function $\xi(m)$ and a period distribution function $f_{lP}$. It is normalised such that
\begin{equation}
 \int\limits_{-\infty}^{\infty}f_{lE}(lE)\,dlE=1.
\end{equation}

For the equal-mass binary population used in the present work a binding energy is chosen from the energy distribution function $f_{lE}$ (Fig.~\ref{flEvergl}). The eccentricities, $e$, are then generated from the thermal eccentricity distribution function, $f(e)=2e$ (e.g. \citealt{du91}).

\end{enumerate}

\subsection{Density profile}
\begin{figure}
      \includegraphics[width=84mm]{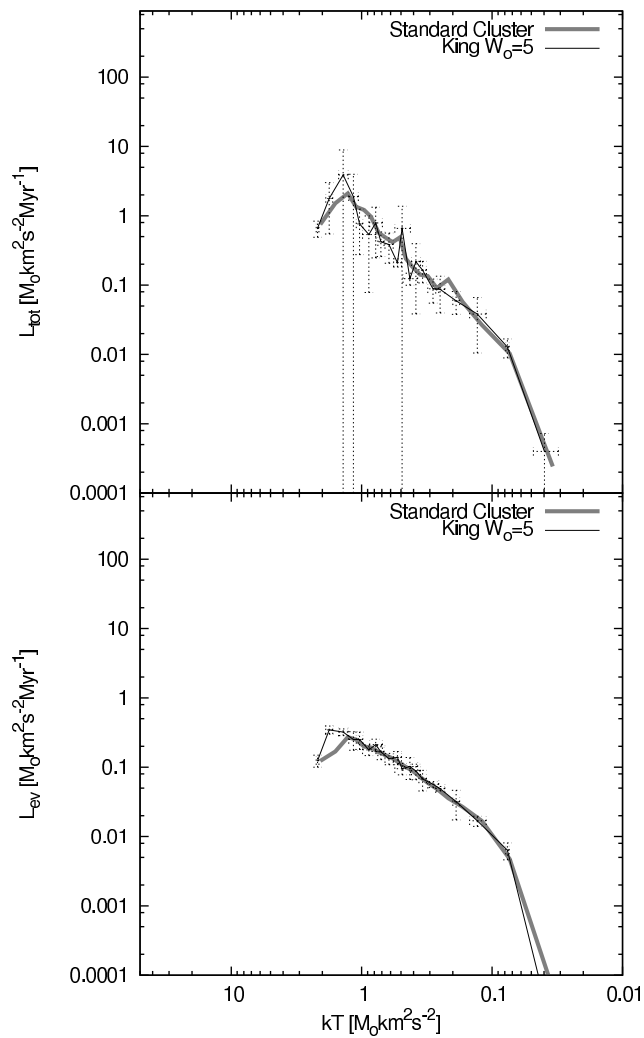}
   \caption{Mean temperature-luminosity diagram of clusters with an initial King density profile ($W_0 = 5$, tidally underfilling by a factor of three). Upper panel: Evaporation + ejection. Lower panel: Evaporation. For comparison the thick grey lines give the cooling track of the \textit{Standard Cluster} (Fig.~\ref{1000tl}).}
  \label{evking}
\end{figure}
Fig.~\ref{evking} depicts that, as expected, the tidally underfilling King models do not differ significantly from the \textit{Standard Cluster}. Within the first few crossing times the initial density profile is completely lost, the King models therefore follow the same cooling track as the \textit{Standard Cluster}. Furthermore, the half-mass radius of those models settles to the same equilibrium value of about 2 pc (see Tab.~\ref{table1}).

\subsection{Half-mass radius}\label{ssec:Rh}
\begin{figure}
    \includegraphics[width=84mm]{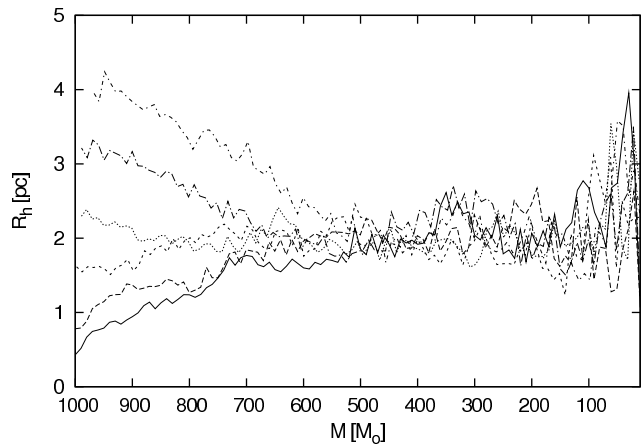}
  \caption{Half-mass radii of models with initial $R_h$ of 0.4, 0.8, 1.6, 2.4, 3.2 and 4.0 pc. All converge to an equilibrium value of 2 pc after core collapse, when the clusters have lost about 30-50\% of their initial mass.}
  \label{rhs}
\end{figure}

\begin{figure}
      \includegraphics[width=84mm]{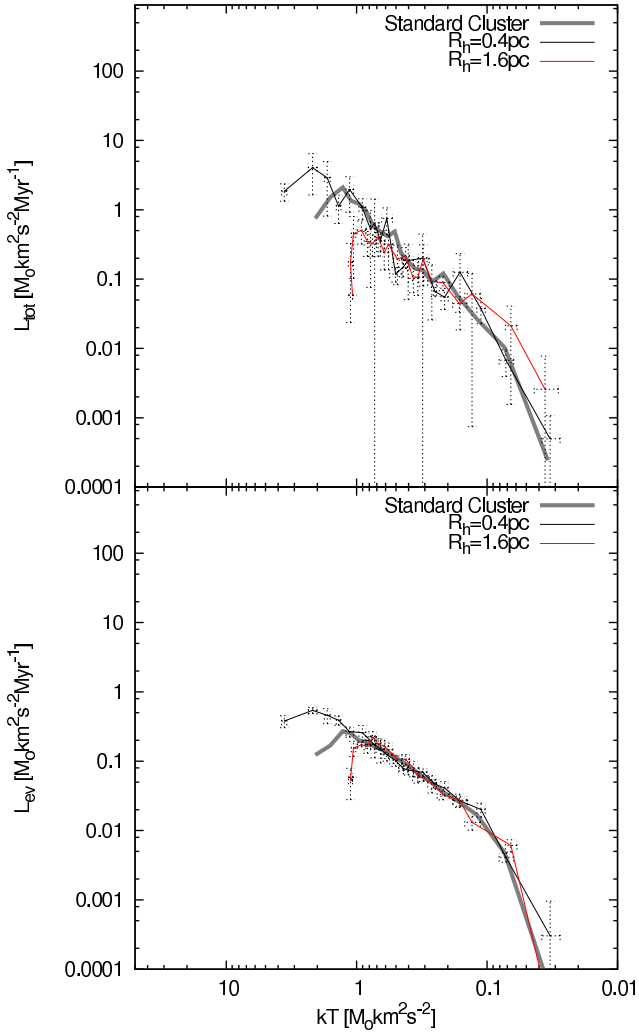}
   \caption{Mean temperature-luminosity diagram of clusters with initial half-mass radii of 0.4 and 1.6 pc. Upper panel: Evaporation + ejection. Lower panel: Evaporation. For comparison the thick grey lines give the cooling track of the \textit{Standard Cluster} (Fig.~\ref{1000tl}).}
  \label{evrh}
\end{figure}
 
\begin{figure}
  \includegraphics[width=84mm]{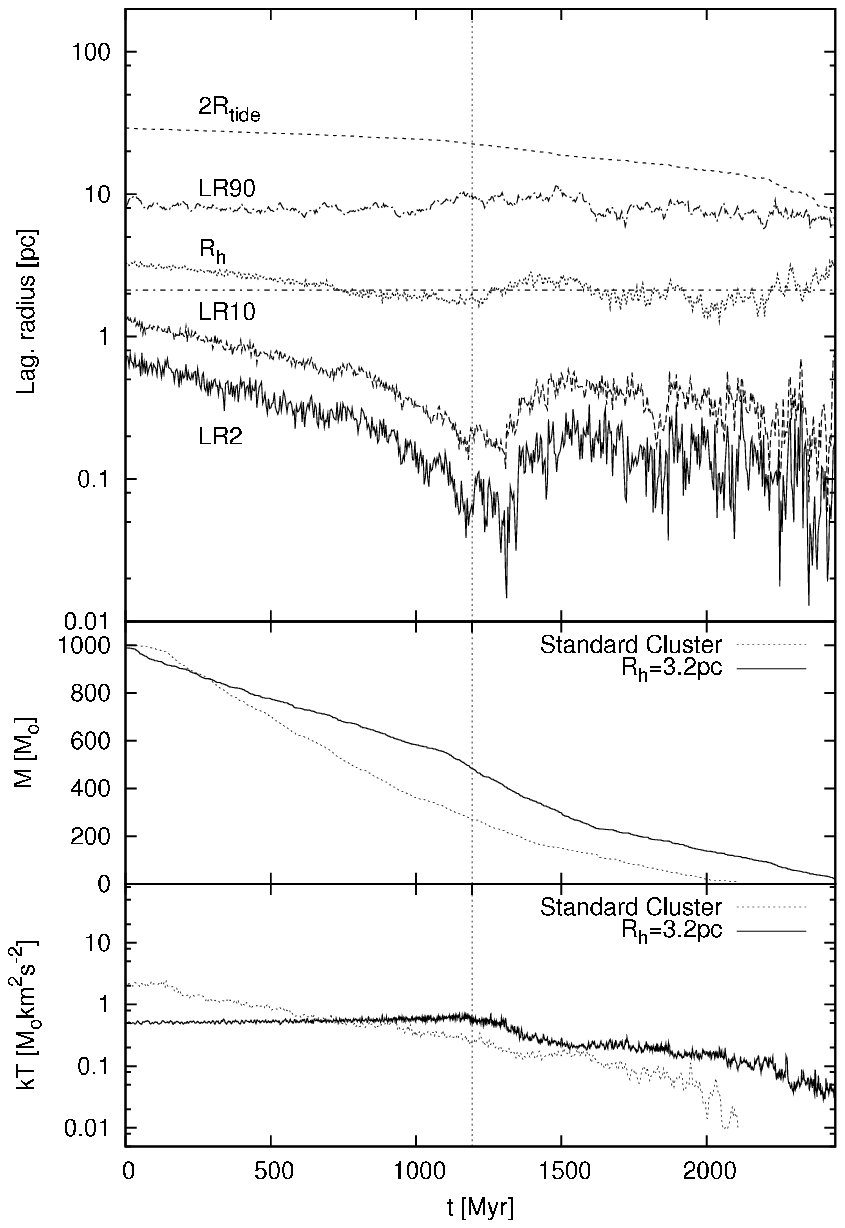}
  \caption{Above: Evolution of the Lagrangian radii for a cluster with initial $R_h=3.2$ pc. Core collapse takes about half the cluster's whole lifetime. The middle panel shows the mass evolution of the same cluster compared to a \textit{Standard Cluster}, which is dissolving more slowly at the very beginning. In the bottom panel the temperature of the cluster is shown. During evolution towards core collapse the temperature stays nearly constant.}
  \label{dMdiff}
\end{figure}

Due to virial equilibrium, reducing the initial half-mass radius raises the initial temperature of the system. A model with half-mass radius of 0.4 pc therefore exhibits a temperature which is twice the temperature of a \textit{Standard Cluster}, while a model with 1.6 pc starts with half of this temperature.

But since these clusters have the same tidal radius, they redistribute their mass through two-body relaxation such that after core collapse the initially smaller or larger clusters do not differ in their properties (e.g. temperature, density profile, binary fraction) from a \textit{Standard Cluster} of the corresponding mass, except for their age. This holds especially for the half-mass radius (Fig.~\ref{rhs}), which settles to the same equilibrium value of 2 pc as the \textit{Standard Cluster} (see Tab.~\ref{table1}). This indicates that the evolution after core collapse is independent of the initial density profile, which can also be nicely seen in the \textit{dynamical T-L diagram}: after starting to ``shine'' they all share a common cooling track (Fig.~\ref{evrh}). 

\begin{figure}
    \includegraphics[width=84mm]{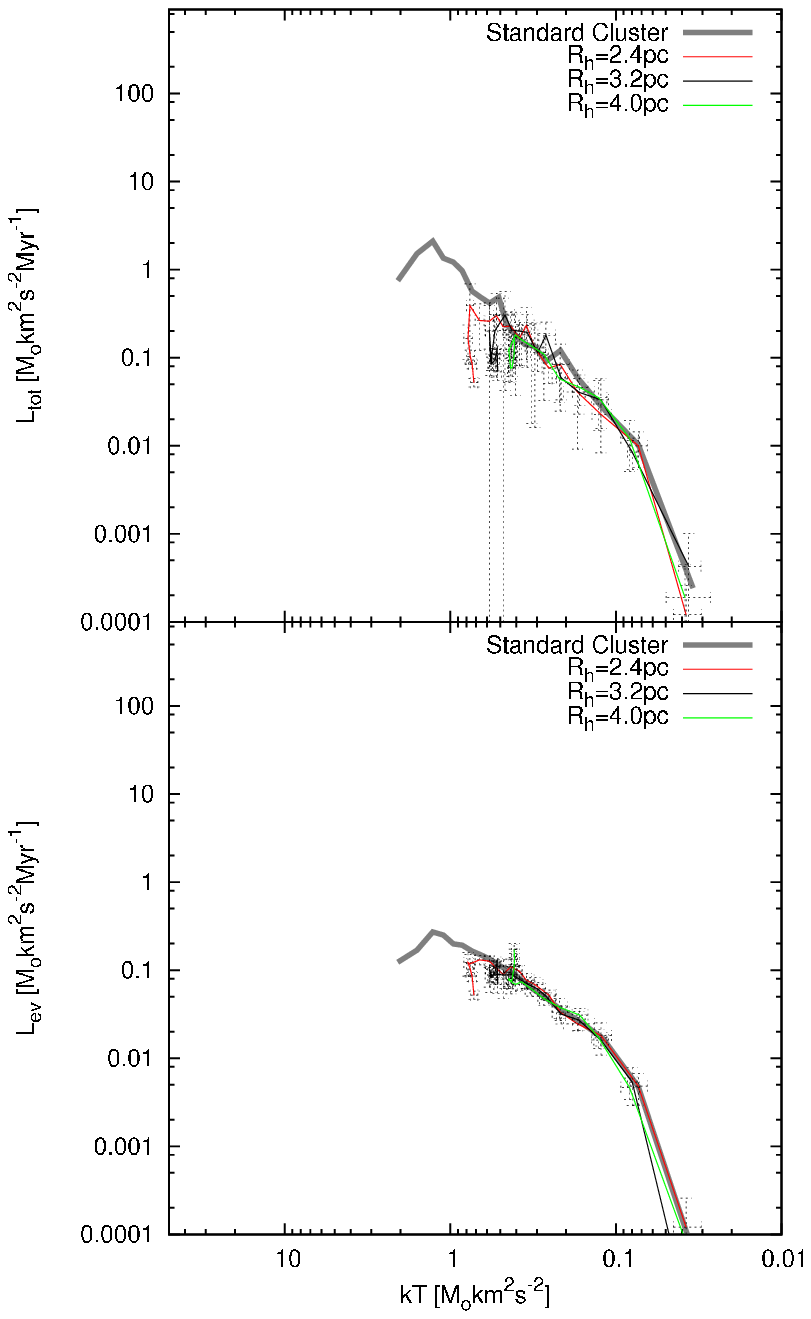}
  \caption{Mean temperature-luminosity diagram of clusters with initial half-mass radii of 2.4, 3.2 and 4.0 pc. Upper panel: Evaporation + ejection. Lower panel: Evaporation. After birth the clusters spend a long time on the same spot along the \textit{dynamical main sequence}. For comparison the grey thick lines give the cooling track of the \textit{Standard Cluster} (Fig.~\ref{1000tl}).}
  \label{ev32}
\end{figure}

A cluster with an initial half-mass radius larger than the equilibrium value of 2 pc, e.g. 3.2 pc, however, evolves very differently to a \textit{Standard Cluster} (Fig.~\ref{dMdiff}). Since the initial median two-body relaxation time, given by
\begin{equation}\label{eq:trel}
 t_{rel} = 0.138 \frac{N^{1/2}R_h^{3/2}}{\overline{m}^{1/2}G^{1/2}\ln\!\Lambda},
\end{equation}
where $\overline{m}$ is the mean stellar mass and $\ln\!\Lambda$ is the Coulomb logarithm \citep{Spitzer87}\footnote[1]{\citet{Gi94} find that the Coulomb logarithm can be approximated by $\ln\!\Lambda \simeq \ln \left(0.11 N\right)$.}, is about ten times larger than for a \textit{Standard Cluster}, evolution towards core collapse will take much longer. In fact, the ratio of core collapse time and intital relaxation time is supposed to be a constant with value about 17 for Plummer models \citep{Ba02}, hence it takes 10 times longer in such models to reach core collapse, while the dissolution time is barely changed (compare with Tab.~\ref{table1}). During this time the half-mass radius will be shrinking until the equilibrium value is established. This has important consequences for the evolution of the other cluster parameters.

The middle panel of Fig.~\ref{dMdiff} shows, that mass loss is more rapid at the very beginning, because the larger cluster already fills the tidal sphere. This effect becomes more important the larger the cluster is. But, as energy production and hence expansion is not so efficient, the \textit{Standard Cluster} dissolves faster in the end.

As in all other models, no binary ejecters are produced before core collapse. The total energy of these models is therefore nearly constant over a large time span, so is the temperature (lower panel of Fig.~\ref{dMdiff}).

This has significant influence on the evolution in the \textit{dynamical temperature-luminosity diagram}. Since the temperature stays constant for about half of the cluster's lifetime, it spends much time at the same position on the common cooling track, until its structure has reached the \textit{Standard Cluster} configuration and starts moving along with it until dissolution (Fig.~\ref{ev32}). This is an excellent proof for the existence of a tight relation between temperature and luminosity. Furthermore this shows that the cooling sequence is also a \textit{dynamical main sequence}, since star clusters with half-mass radii larger than the equilibrium value stay at the same spot on the sequence until their internal configuration has adopted to the tidal conditions, which can take up to half of their total life-time. 

The clusters with an initial $R_h$ of 4.0 pc lose so many of their stars in the very beginning due to their initial dimension, that their luminosity is initially slightly above the \textit{dynamical main sequence}. More extended clusters would have a even higher initial luminosity before they finally reach the \textit{dynamical main sequence}.

\subsection{Number of stars}

\begin{figure}
    \includegraphics[width=84mm]{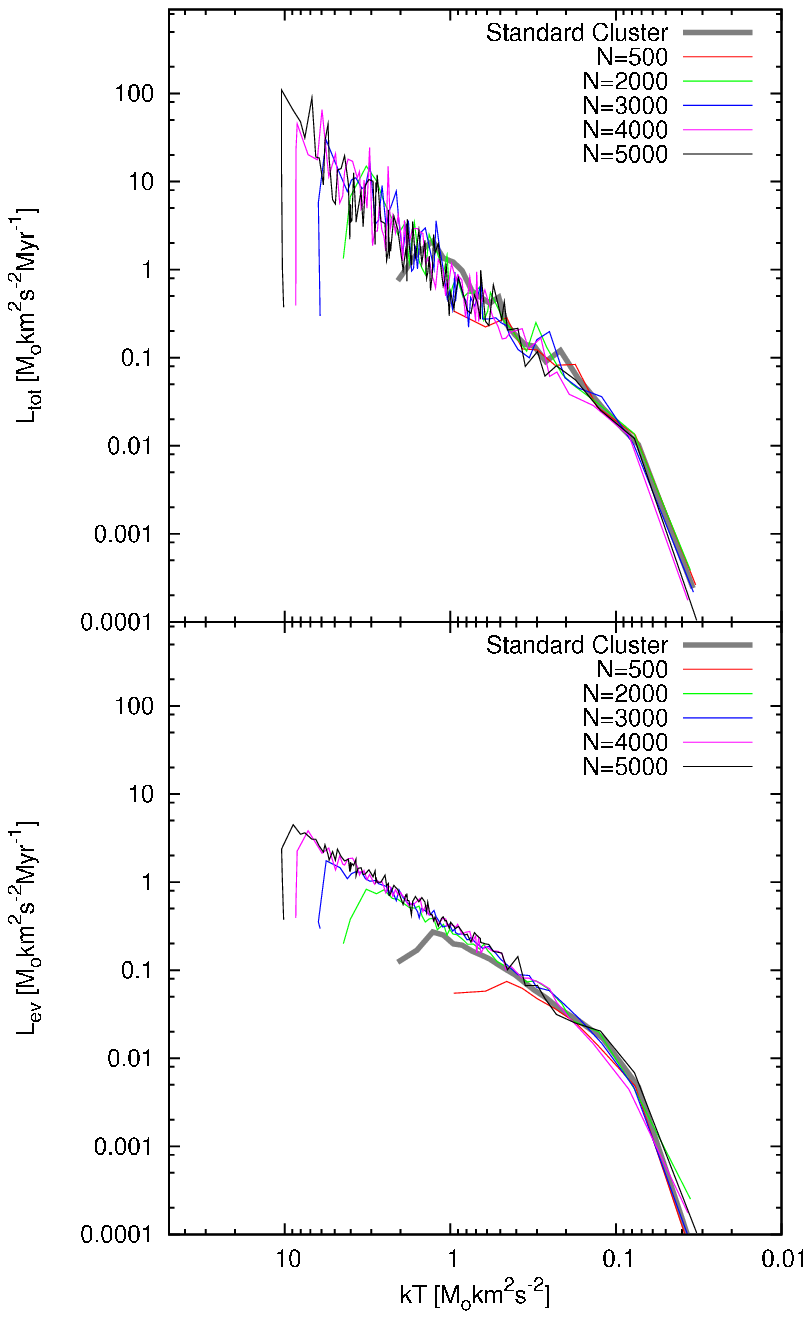}
  \caption{Temperature-luminosity diagram for clusters of different initial masses. Upper panel: Evaporation + ejection. Lower panel: Evaporation. Each cluster starts on the left and moves rapidly onto the common cooling sequence. For comparison the thick grey lines give the cooling track of the \textit{Standard Cluster} (Fig.~\ref{1000tl}).}
  \label{cseq}
\end{figure}
\begin{figure}
    \includegraphics[width=84mm]{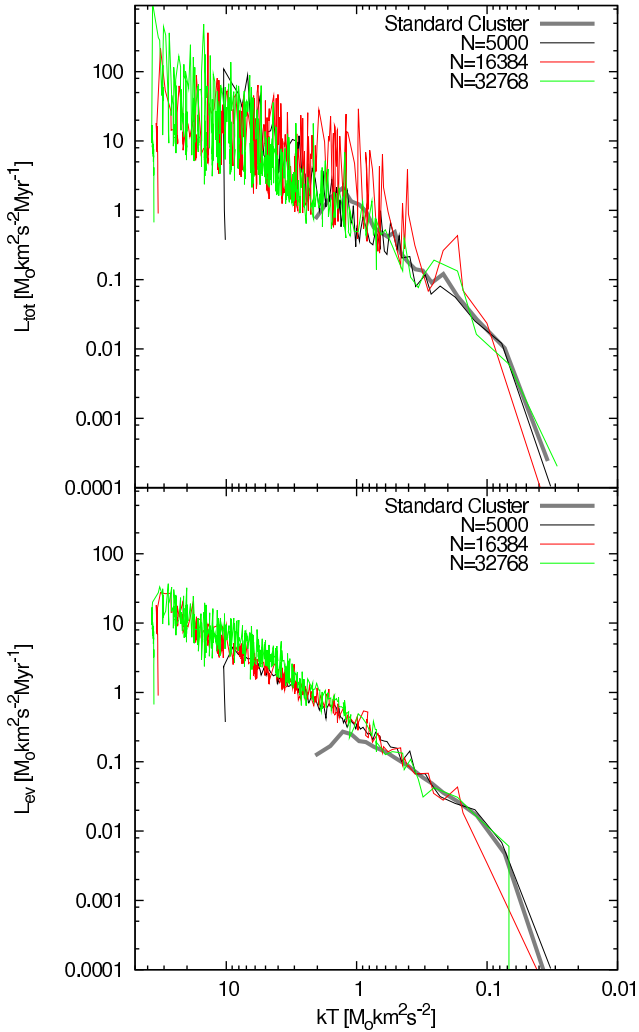}
  \caption{Temperature-luminosity diagram for the two clusters with masses of 16384 and 32768 $\msun$ and the clusters with 5000 $\msun$. Upper panel: Evaporation + ejection. Lower panel: Evaporation. Each cluster starts on the left and moves rapidly onto the common cooling sequence. For comparison the thick grey lines give the cooling track of the \textit{Standard Cluster} (Fig.~\ref{1000tl}).}
  \label{cseqlarge}
\end{figure}

\begin{figure}
    \includegraphics[width=84mm]{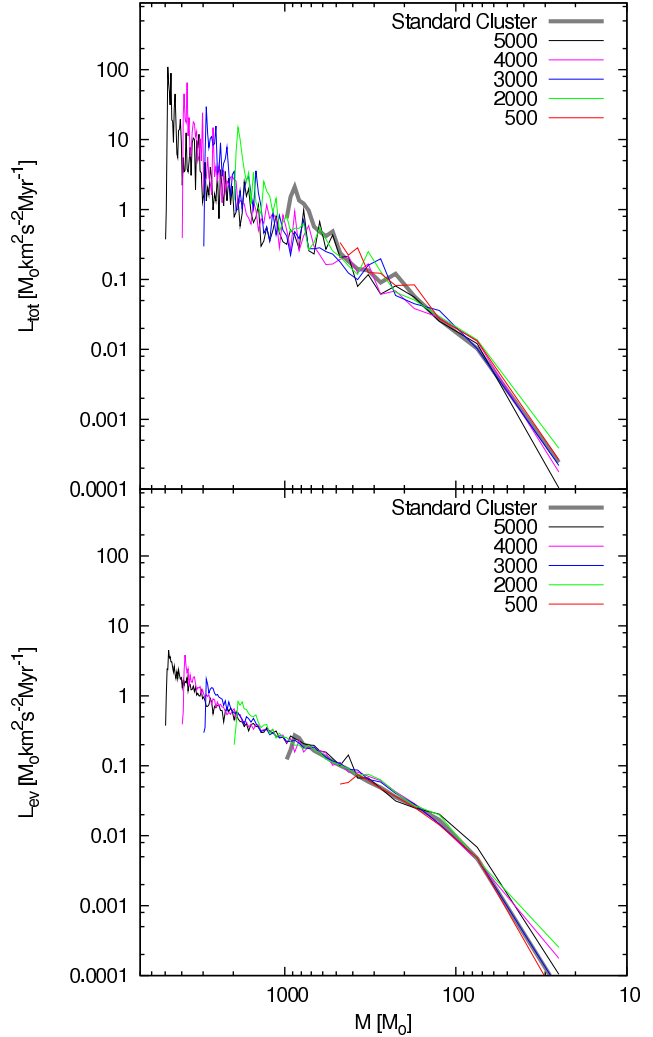}
  \caption{The same as Fig.~\ref{cseq} but with the cluster mass instead of the temperature on the x-axis. After core collapse all models are fully determined by their mass. Upper panel: Evaporation + ejection. Lower panel: Evaporation. For comparison the thick grey lines give the evolution of the \textit{Standard Cluster}.}
  \label{cseqm}
\end{figure}

\begin{figure}
    \includegraphics[width=84mm]{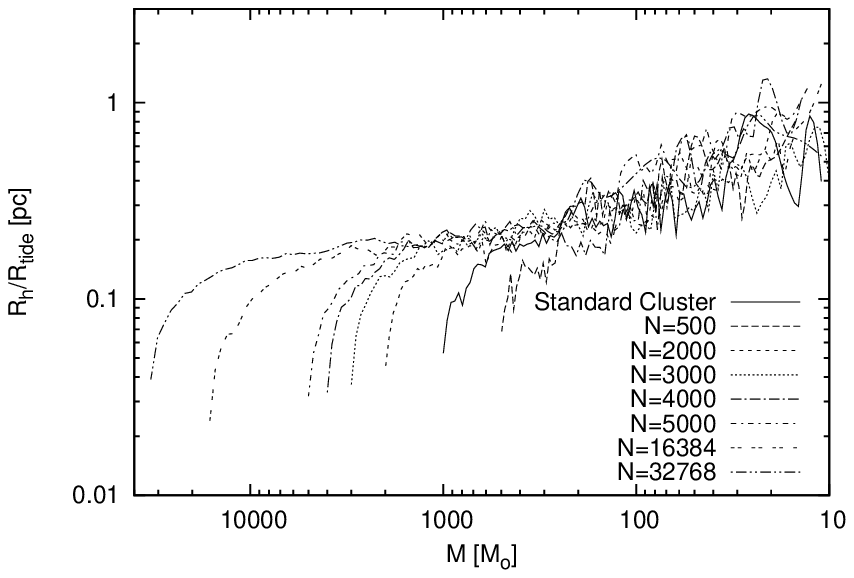}
  \caption{Ratio of half-mass radius to tidal radius for clusters with masses of 500 to 32768 $\msun$. The clusters asymptotically evolve onto a common sequence. Below 100 $\msun$ fluctuations in $R_h$ lead to a large scatter around this sequence.}
  \label{rhrtsn}
\end{figure}

Increasing the initial mass of the cluster causes a deeper potential well and hence a higher temperature. However, all clusters evolve along a common cooling sequence independently of their initial masses (Fig.~\ref{cseq}). In Fig.~\ref{cseqlarge} the results of the most massive clusters are shown. This was done seperately for clarity reasons because for $N=32768$ and $N=16384$ only one model each was computed, which results in a larger scatter. Nevertheless the two clusters clearly extend the cooling sequence to higher temperatures.
Investigating the luminosity in dependence of the mass shows that after core collapse all models evolve on a common track, hence are fully characterised by the mass left in the cluster (Fig.~\ref{cseqm}).

This is also emphasized in Fig.~\ref{rhrtsn}, which shows the ratio of half-mass radius to tidal radius as a function of the mass left in the cluster. It can be seen that all models start tidally underfilling and increase their $R_h/R_{tide}$ initially, after which they evolve asymptotically onto a common sequence, independent of their starting condition.
On this sequence the ratio $R_h/R_{tide}$ increases, which could be interpreted in the way that, as the half-mass relaxation time (eq. \ref{eq:trel}) decreases with decreasing mass, energy production within the half-mass radius gets more efficient, leading to a nearly constant equilibrium half-mass radius opposed to a declining tidal radius.

This supports the assumption that, after core collapse, the state of a cluster in a given tidal condition just depends on its mass. All initial conditions are lost through core collapse and a universal density distribution is established.

This finding of a constant $R_h$ shows that the simplified theory of cluster evolution set-up for illustrative purposes by \citet[p.~525]{Bi87} does not capture cluster evolution correctly. According to this ansatz the half-mass radius decreases with ongoing mass loss, as
\begin{equation}
 R_h(t)=R_h^0\left(\frac{M(t)}{M_0}\right)^2,
\end{equation}
where $R_h^0$ and $M_0$ are the initial values of the half-mass radius and the cluster mass. In this ansatz it is assumed, that evaporation is the dominant mode of escape which leaves the energy of the cluster nearly constant, as \textit{evaporaters} take away almost no energy. This holds true for initially very extended clusters for the time before core collapse, since no \textit{ejecters} have been produced yet. If the cluster is too concentrated in the beginning, i.e. the initial half-mass radius is smaller than the equilibrium value, the cluster expands while the core is contracting and therefore the ansatz also does not hold.

Furthermore, the finding that $R_h/R_{tide}$ increases also contradicts the prediction of \citet[p.~59]{Spitzer87}, who found the ratio of $R_h/R_{tide}$ to be a constant of about 3, i.e. the cluster to evolve self-similarly. Instead, this ratio is a function of mass, and the value of 3 is reached just at the very end of a cluster's life-time.

For the underlying set of investigated models, the constant value of the half-mass radius is a much better and more useful approximation. Therefore the half-mass radii of all clusters have been studied in detail and a constant has been fitted to $R_h$ of each cluster between $t_{cc}$ and the dissolution time, $t_{dis}$. In Tab.~\ref{table1} these fitted mean values, $R_h^f$, for the half-mass radii of all model classes are shown.

\subsection{Tidal field}\label{sec:tf}
\begin{figure}
  \includegraphics[width=84mm]{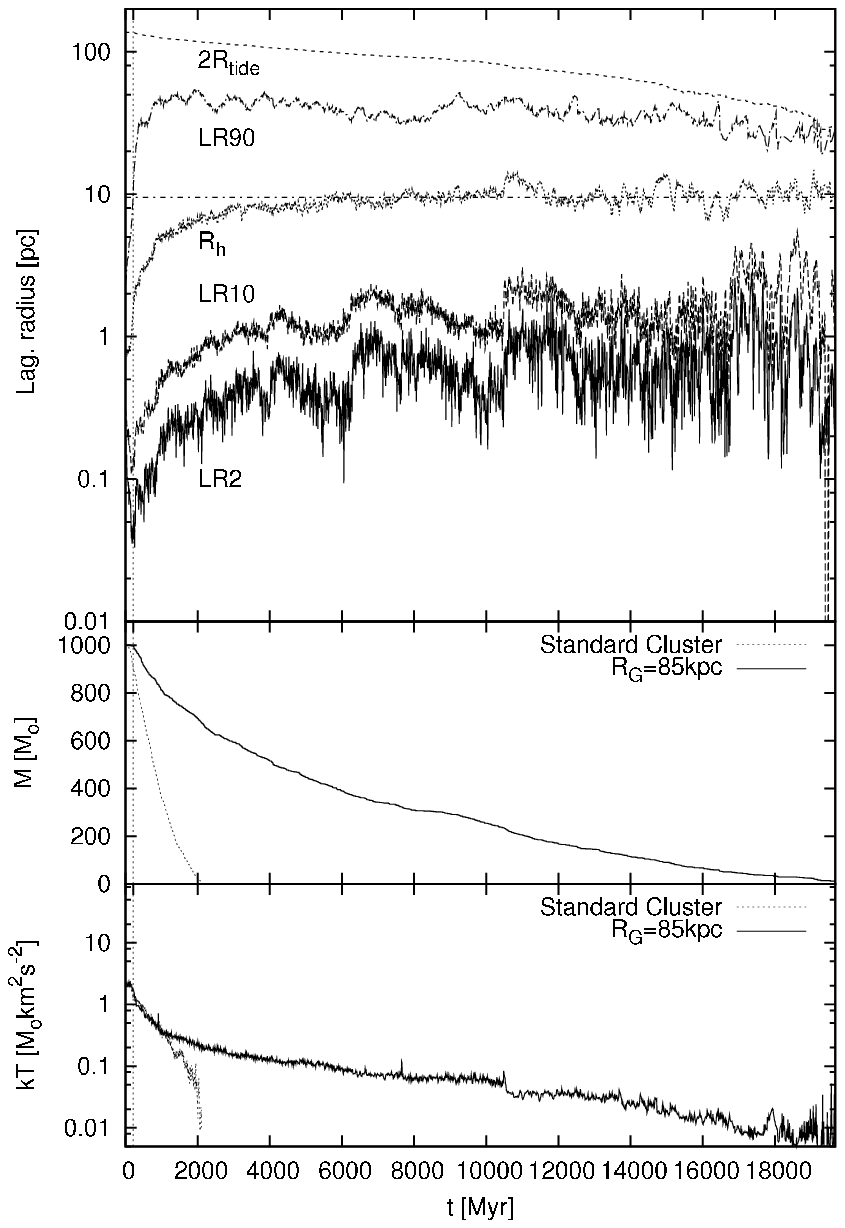}
  \caption{Above: Evolution of the Lagrangian radii for a cluster with $R_G=85$ kpc. After core collapse the cluster is still expanding until it fills the tidal sphere at about 2000 Myr. The middle panel shows the mass evolution of the same cluster compared to a \textit{Standard Cluster}, which is much slower due to the larger tidal radius and the increasing dynamical time-scale. In the bottom panel the temperature of the cluster is shown, which differs from the \textit{Standard Cluster} after the latter has filled its tidal sphere, while the former continues expanding.}
  \label{dMdiffrgal}
\end{figure}

\begin{figure}
    \includegraphics[width=84mm]{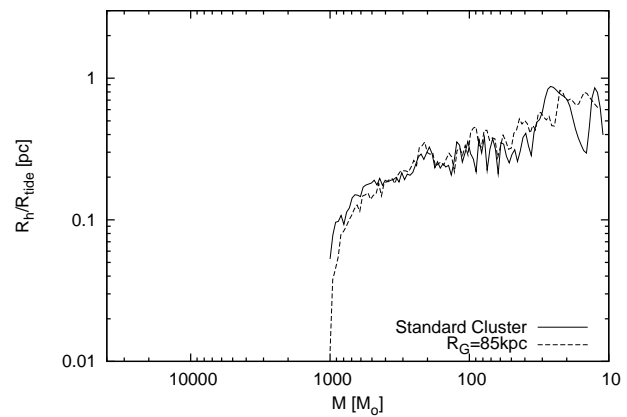}
  \caption{Ratio of half-mass radius to tidal radius for a cluster at Galactocentric distance of 85 kpc compared to a \textit{Standard Cluster} at 8.5 kpc.}
  \label{rhrtsrg}
\end{figure}

\begin{figure}
    \includegraphics[width=84mm]{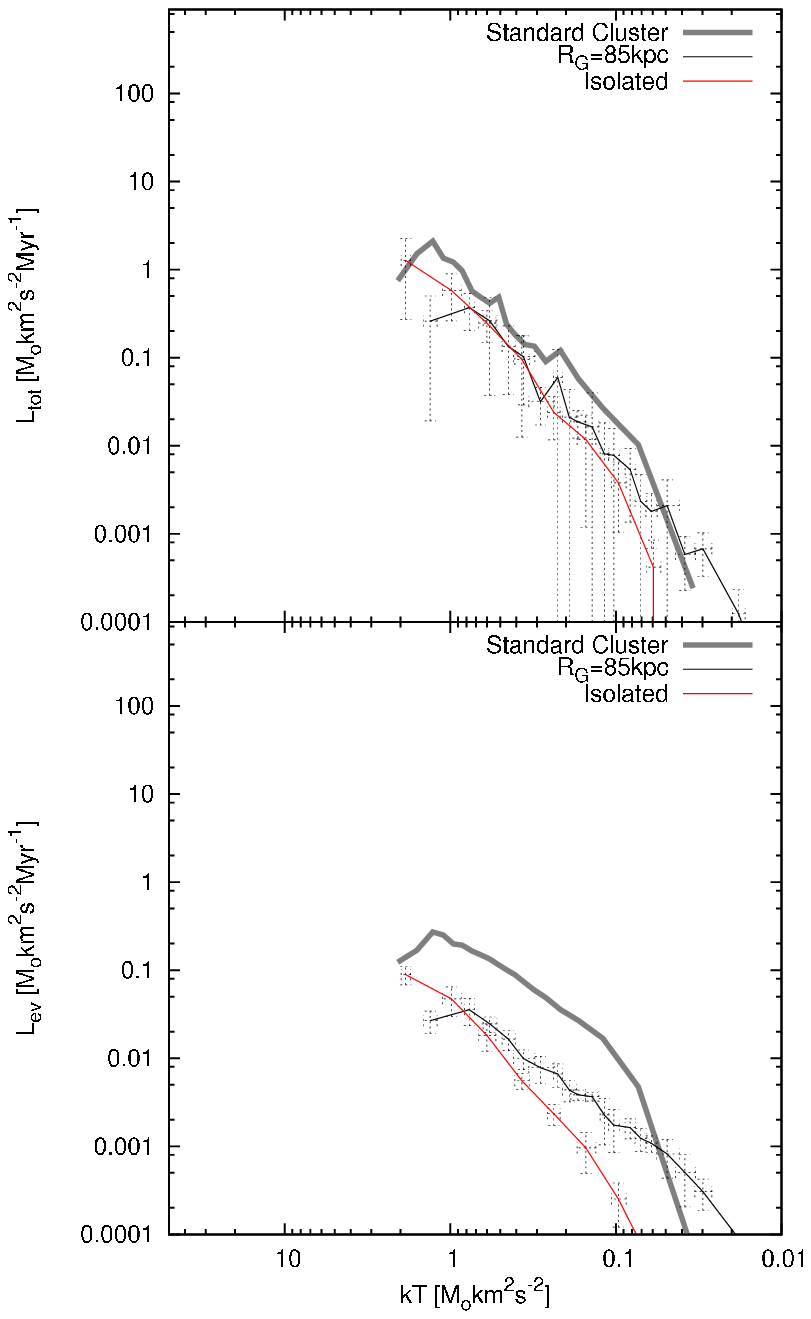}
  \caption{Mean temperature-luminosity diagram of an isolated cluster and one with a Galactocentric distance of 85 kpc. Upper panel: Evaporation + ejection. Lower panel: Evaporation. For comparison the thick grey lines give the cooling track of the \textit{Standard Cluster} (Fig.~\ref{1000tl}).}
  \label{evrg}
\end{figure}

The Galactocentric distance was set to 85 kpc for a number of models. Here the tidal force is much weaker and the tidal radius (eq. \ref{eq:rtide}) is about five times larger, which explains the following observed differences, since these are just due to a different scaling.

Initially the cluster is almost a \textit{Standard Cluster} with the same structure and relaxation time. It goes into core collapse on the same timescale, but the outer shells expand, due to the weaker tidal field, much further than those of a \textit{Standard Cluster} (Fig.~\ref{dMdiffrgal}). This increases the dynamical time-scale of the cluster. After about 2 Gyr, the cluster has filled its tidal sphere and the half-mass radius then reaches a constant state. This shows that the tidal conditions have a strong influence on the equilibrium value of the half-mass radius. But taking a look at Fig.~\ref{rhrtsrg} reveals that, after the equilibrium density distribution is established, the ratio of half-mass radius to tidal radius is the same for both clusters.

Since encounters of stars are less frequent due to the smaller cluster density, and hence a lower temperature (Fig.~\ref{dMdiffrgal}), the luminosity is much smaller than for a  \textit{Standard Cluster} (Fig.~\ref{evrg}). The cooling tracks are therefore shifted within the T-L diagram. But in $N$-body units the luminosity curves of the two cluster types would almost lie on top of each other, since after core collapse (or more precisely in this case: after both clusters have filled their tidal spheres) they just differ in terms of scaling. For example the total number of binary ejecters, which is independent of scaling, is therefore comparable to a \textit{Standard Cluster} (also about 10).

The influence of a tidal field on cluster evolution can be observed in a direct comparison of a \textit{Standard Cluster} with an isolated cluster of the same properties. Without a tidal field there is no tidal radius and therefore the criterion of escape has to be modified. Here, an escaper is defined as a star with positive energy, since this is the only way for a star to escape from an isolated cluster. The velocities of the escaping stars are not measured until they reach the 90\% Lagrange radius. Otherwise the influence of the cluster potential on the stars' velocities would be too significant. Furthermore, scattering back to bound energies is very unlikely after reaching this radius.

This difference in the escape criterion has some effect on the evolution in the T-L diagram (Fig.~\ref{evrg}). Since the 90\% Lagrange radius is, during the first Gyr, much smaller than 2 $R_{tide}$ for the clusters at 85 kpc, the initial luminosity is above the level of those clusters because stars can fulfill the escape criterion more readily. But as cluster expansion goes on forever in an isolated cluster, due to the missing tidal field, escape is getting harder, since the 90\% Lagrange radius also grows inexorably -- and so does the half-mass radius. 

A tidal field therefore counteracts the cluster expansion, by removing the outer stars. An equilibrium value for the half-mass radius is established after core collapse, which guarantees a balance between energy production in the core and energy loss through the tidal field. The absence of such a tidal field prevents the removal of weakly bound stars, so the cluster can expand further and further. 

Hence, after core collapse not only the number of stars left in the cluster are important but also the tidal conditions.

\subsection{Binaries}\label{ssec:bin}
\begin{figure}
    \includegraphics[width=84mm]{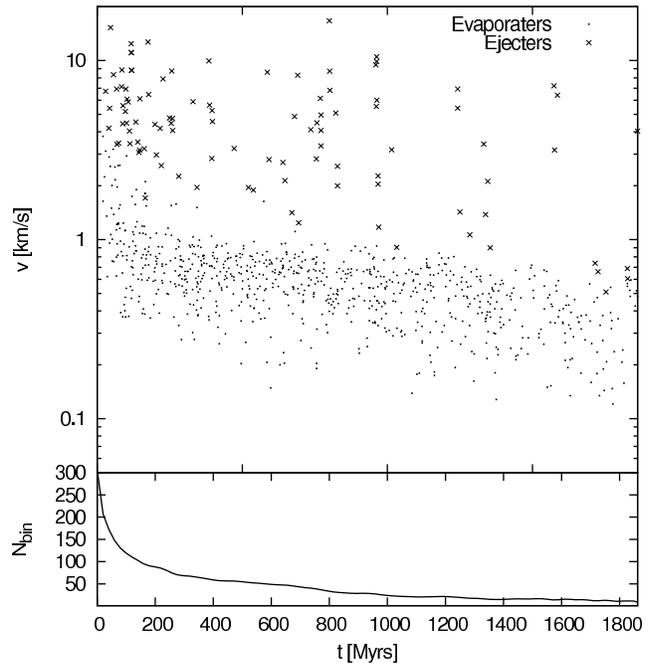}
  \caption{Upper panel: velocities of escaping stars as a function of time for a cluster with $f_{bin}=0.95$. Below: number of binaries over the course of time. Most primordial binaries are burnt to prevent the core from collapsing and drive the halo expansion.}
  \label{vtbin}
\end{figure}

\begin{figure}
 \includegraphics[width=84mm]{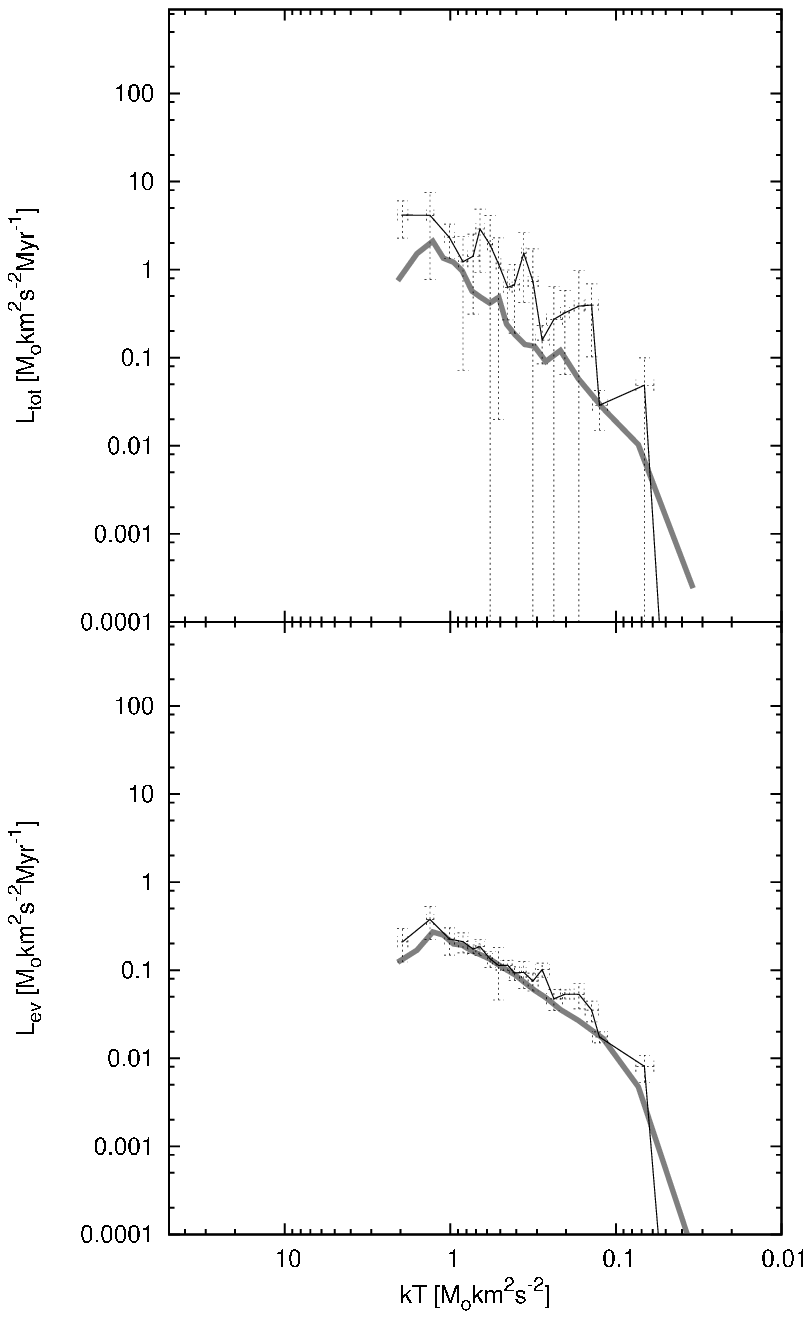}
 \caption{Mean temperature-luminosity diagram of the clusters with 95\% primordial binaries. Upper panel: Evaporation + ejection. Lower panel: Evaporation. For comparison the grey lines give the evolution of the \textit{Standard Cluster} (Fig.~\ref{1000tl}).}
 \label{evfb}
\end{figure}

Effects of primordial binaries on star cluster evolution have been studied thoroughly \citep*{Heggie75,Mc90,Kr95c}. In this investigation models with $f_{bin}=0.95$ are set up with 5\% single stars and 95\% centre of mass particles first. Thereafter the c.m. particles are substituted with a random binary, following the binding energy distribution $f_{lE}$ (eq.~\ref{flE}). 

Like in a \textit{Standard Cluster} the initial temperature of single stars and c.m. particles is about $kT = 2\, \msun\mbox{km}^2\mbox{s}^{-2}$. So, assuming that binaries with $E_{bin}$ less than $kT$ are disrupted immediately, setting up a binary fraction of $f_{bin} = 0.95$ with this binding energy distribution yields about 300 bound binary systems initially in the cluster (Fig.~\ref{vtbin}).

One major consequence of the presence of a large number of primordial binaries is the well known effect of a damped core collapse. With a lot of energy sources available, the cluster has no need to go into deep collapse to produce energy (see \citealt{Ostriker85}). Furthermore, binary-binary interactions are much more efficient in energy production and therefore allow a lower core density \citep{Hut96}. Another way to put this is an analogon to a protostar which has a significant amount of primordial deuterium. Since deuterium burning is efficient at lower temperatures compared to hydrogen burning, the whole star can achieve stability at a larger radius.

This effect is accelerated by the fact that, since binary systems in single-mass clusters are subject to mass segregation, they sink to the cluster core within a few crossing times. So most binaries are burnt in the core within a few relaxation times at the very beginning of the computations and just a few remain (Fig.~\ref{vtbin}).

The presence of primordial binaries also increases the relative number of \textit{ejecters}, since there are many binaries to produce them already from the start (upper panel in Fig.~\ref{vtbin}). Hence, the \textit{ejection luminosity} is much larger than in a \textit{Standard Cluster} while the evaporation is unaffected by this (Fig.~\ref{evfb}). Nevertheless the limit between \textit{evaporaters} and \textit{ejecters} is the same as in a \textit{Standard Cluster}.

Furthermore, the first \textit{ejecter} cannot be taken as a trigger for core collapse, contrary to the case of single-star clusters. But here also the standard criterion for core collapse of the first binary with a binding energy above a certain value naturally fails.

Since most of the primordial binaries are disrupted in binary-binary encounters and just a few are ejected, much of their kinetic energy gets distributed among the other stars. This constant heat source leads to a more extended cluster and hence a different half-mass radius (see Tab.~\ref{table1}). So after core collapse, a cluster may not be fully described by only the mass and the tidal conditions, but also by the binary fraction.

\section{Conclusions}
The investigations made here show that it is possible to define a \textit{dynamical temperature} and a \textit{dynamical luminosity} for star clusters. The \textit{dynamical temperature-luminosity diagram} established with these two quantities gives a completely new way of looking at cluster evolution and helps to gain valuable insights on the energy evolution of star clusters.

The \textit{dynamical temperature} is defined through the mean kinetic energy of the stars within a cluster, where binary systems need to be replaced by their centre-of-mass particles. Through this correction the system can still be treated as an ideal gas, which enables to relate the temperature to the velocity dispersion of the stars within the cluster. In this way the \textit{dynamical temperature} is directly correlated to the velocity distribution of the cluster stars, which is taken to be Maxwellian shaped. The latter implies that there is always a high-velocity tail of stars with velocities above the escape velocity. When these unbound stars leave the cluster, the left-over stars will reestablish a velocity dispersion with a lower \textit{dynamical temperature} within a relaxation time. A relation between \textit{dynamical temperature} and the number of escaping stars is therefore expected, since there will always be a certain fraction of stars leaving the cluster.

The \textit{dynamical luminosity} is defined as the kinetic energy of the stars leaving the cluster, in analogy to the energy of photons emitted by a star. Escaping stars show two different origins, evaporation and ejection, which have to be treated separately. The former is strongly correlated to the temperature, while the latter is due to binary interactions and hence a tracer for structural effects like core collapse or core oscillations. In fact, there is no ejection before core collapse and the occurrence of the first ejected star can be used to define the point in time when the core has reached its densest phase. The limit between \textit{evaporaters} and \textit{ejecters} was found to be universal among the investigated clusters at $E_{kin}/(kT) = 5.1 \pm 0.2$. These two families of escapers are furthermore the reason for splitting up the total luminosity into an \textit{evaporation luminosity} and an \textit{ejection luminosity}. 

In this way the energy evolution of a star cluster can be understood as follows: A cluster is generating energy in the core through binary burning which causes expansion of the whole cluster. Without a tidal field this leads to infinite expansion, since a single binary can generate large amounts of energy until it is ejected from the cluster, in which case the core has to contract to form a new one. 
Stars leaving the cluster have to have positive energy, which they can gain through two-body relaxation processes. This resulting evaporation of stars decreases the total energy of the cluster. Ejection mechanisms have to counteract this decrease of energy. Three- or more-body encounters in the core cause not only the ejection of stars with high velocities, but also the ejection of binaries. The latter therefore dissipate the accumulated binding energy and make the total energy of the cluster go to zero (see Fig.~\ref{E}). A tidal field is setting a limit to the expansion of the cluster, such that the cluster spills over the tidal radius. Evaporation is therefore amplified, since escape is eased.

All computed clusters show a tight relation between \textit{evaporation luminosity} and \textit{dynamical temperature}, while the \textit{ejection luminosity} shows a much larger scatter due to the small number of \textit{ejecters}, i.e. insufficient statistics. Before core collapse all clusters move towards a \textit{dynamical main sequence}, where they spend up to half of their total life-time. After core collapse they follow a common cooling sequence, similar to a cooling track of white dwarfs. The duration of the main sequence phase depends on the initial density distribution of the cluster. For initially very extended clusters, i.e. with large initial $R_h/R_{tide}$, core collapse takes a large fraction of the cluster's total life time, during which the temperature stays nearly constant, corresponding to a fixed position in the \textit{dynamical temperature-luminosity diagram}.

This leads to a large degeneracy in models for equal-mass clusters, since the initial conditions of a cluster on the cooling sequence cannot be traced back. For the \textit{evaporation luminosity} the only deviations from this cooling sequence are given by models with different tidal conditions, i.e. at different Galactic radii, while the \textit{ejection luminosity} is additionally affected by the binary fraction. On the other hand this supports the hypothesis that after core collapse, the state of a single-mass system can be fully described by three quantities: the number of stars left in the cluster, the tidal conditions and the binary content. A theory of cluster evolution therefore has to focus on these parameters.

Whether clusters with a stellar mass distribution would also form a \textit{dynamical main sequence} still has to be investigated. If this is the case, well observed clusters like the Pleiades may be placed within such a diagram and a unique evolutionary track may be assigned to them, which would give direct insights on the former and further evolution of those systems.

Another important finding of this parameter-space study is the phenomenon of a constant half-mass radius for clusters in tidal fields. This holds for the time after core collapse, when a final density distribution is established within the tidal radius, in which the energy production in the core is balanced by the mass loss at the tidal radius. As shown in this work, the tidal field has a significant influence on this value, since clusters at larger Galactic radii show a much larger equilibrium value for the half-mass radius, while isolated clusters do not show this phenomenon at all. This can be expressed in terms of the ratio of half-mass radius to tidal radius, which evolves along a common sequence for all investigated clusters, depending only on mass. The ratio is increasing with time, as the half-mass radius is a constant and the tidal radius is decreasing due to ongoing mass loss. This implies that the energy production efficiency within the half-mass radius is increasing as the cluster loses mass.

The \textit{Standard Cluster} shows an equilibrium half-mass radius of about 2 pc in the given tidal field, and so do the other models with the same initial mass, tidal conditions and binary fractions (Tab.~\ref{table1}).
An increasing initial mass increases the value of the equilibrium half-mass radius. The models with a higher mass than the \textit{Standard Cluster} have a slightly declining half-mass radius at the end of their lifetimes, when their ratio of $R_h/R_{tide}$ has reached the common sequence. While the model with 16384 stars still shows a nearly constant $R_h$, the 32k model has a slightly increasing half-mass radius after core collapse until about half its dissolution time, after which it starts decreasing slowly. In a first order approximation it still can be described by a constant but it is expected that this gets more and more inexact the larger the initial mass of the cluster is.

This behaviour contradicts the theory of self-similar evolution of \citet[p.~59]{Spitzer87}, who found the ratio $R_h/R_{tide}$ to be a constant of about 3, which seems to be a too rough estimate and just holds at the very end of a cluster's life-time. Furthermore, the simple cluster-evolution theory based on the ansatz that evaporation does not change the energy of a cluster and thus the half-mass radius scales with $M^2$ \citep[p.~525]{Bi87}, can be completely ruled out. This only holds for very extended clusters, for which the initial half-mass radius is larger than the equilibrium value, and for these clusters this ansatz holds only until core collapse because until then no ejection has occurred. In this given set of parameters the constant half-mass radius gives a much better approximation. Nevertheless, this parameter-space study showed how similar single-mass clusters with a wide range of initial conditions evolve, once they have adjusted to the given tidal conditions, which is in most cases right after core collapse.

The discovery of Russell and Hertzsprung was made possible through improvements in distance-determination techniques for stars, just like future developments like GAIA will increase the possibilities of measuring accurate peculiar velocities of stars in the Milky Way. Then, also a large sample of open clusters will be surveyed in detail, which will give the opportunity to reconstruct internal quantities like the velocity dispersion very accurately and also to identify stars which are about to leave a cluster or have left it lately. This means that the actual measurement of the two quantities, \textit{dynamical temperature} and \textit{dynamical luminosity}, will become feasible.

\begin{table*}
\begin{minipage}{158mm}
\centering
 \caption{Overview of all computed models. $n$ gives the number of computed models of a particular kind, $\rho(r)$ is the initial density profile (P: Plummer, K: King) and $R^f_h$ is the fitted value for the half-mass radius after core collapse. Given errors are the standard deviations. The last column gives the fitted values for the limit between \textit{evaporaters} and \textit{ejecters} with fitting uncertainties.}
\label{table1}
\begin{tabular}{ccccccccccccc}
\hline
  $n$ & $N$& $\rho(r)$  & $R_h$ [pc]  & $f_{bin}$ & $R_{gal}$ [kpc]  & $t_{cc}$ [Myr] & $t_{rel}^0$ [Myr]&$t_{dis}$ [Myr]& $R^f_h$ [pc]& $\left(\frac{E_{kin}}{kT}\right)_{limit}$\\
\hline
27&1000&P&0.8&0&8.5&165$\pm$20&10.1&2277$\pm$117&2.01$\pm$0.09&5.5$^{+1.0}_{-0.9}$\\
\hline
5&1000&K&0.8&0&8.5&162$\pm$37&10.1&2198$\pm$120&1.86$\pm$0.04&5.4$^{+0.8}_{-0.7}$\\
\hline
5&1000&P&0.4&0&8.5&62$\pm$5&3.6&2039$\pm$85&1.96$\pm$0.12&5.7$^{+1.3}_{-1.1}$\\
5&1000&P&1.6&0&8.5&432$\pm$50&28.7&2467$\pm$159&2.15$\pm$0.09&4.8$^{+0.8}_{-0.7}$\\
9&1000&P&2.4&0&8.5&758$\pm$47&52.6&2573$\pm$175&2.18$\pm$0.17& 7.3$^{+8.8}_{-4.0}$ \\
5&1000&P&3.2&0&8.5&1081$\pm$98&81.0&2592$\pm$173&2.09$\pm$0.09&5.9$^{+3.5}_{-2.2}$\\
10&1000&P&4.0&0&8.5&1061$\pm$116&113.3&2489$\pm$94&2.03$\pm$0.13&4.3$^{+1.1}_{-0.9}$\\
\hline
33&500&P&0.8&0&8.5&134$\pm$32&8.4&1662$\pm$174&1.88$\pm$0.12&6.7$^{+1.4}_{-1.1}$\\
8&2000&P&0.8&0&8.5&191$\pm$27&12.5&3195$\pm$150&2.13$\pm$0.05&4.8$^{+0.9}_{-0.8}$\\
5&3000&P&0.8&0&8.5&241$\pm$11&14.2&3871$\pm$101&2.29$\pm$0.03&4.5$^{+1.3}_{-1.0}$\\
4&4000&P&0.8&0&8.5&253$\pm$19&15.6&4563$\pm$316&2.38$\pm$0.14&4.3$^{+1.1}_{-0.9}$\\
3&5000&P&0.8&0&8.5&290$\pm$9&16.9&4943$\pm$180&2.48$\pm$0.13&4.4$^{+1.7}_{-1.2}$\\
1&16384&P&0.8&0&8.5& 442 & 25.7 &  10077 &  2.99 & 3.8$^{+1.6}_{-1.1}$\\
1&32768&P&0.8&0&8.5& 1093 & 33.3 & 15645 &  3.57  & 3.6$^{3.2}_{-1.7}$\\
\hline
5&1000&P&0.8&0&85&193$\pm$19 &10.1&$>$20000&8.69$\pm$0.81&5.7$^{+0.8}_{-0.7}$\\
5&1000&P&0.8&0&$\infty$&146$\pm$31&10.1&$>$20000&(--)\footnote[1]{Isolated clusters do not show a constant half-mass radius.}&4.3$^{+0.6}_{-0.5}$\\
\hline
4&1000&P&0.8&0.95&8.5& (--)\footnote[2]{The presence of primordial binaries prevents the core from collapsing.} &7.8&2012$\pm$232&2.28$\pm$0.14&5.7$^{+0.7}_{-0.6}$\\
\hline
\end{tabular}
\end{minipage}
\end{table*}

\section*{Acknowledgments}
AHWK would like to thank Douglas Heggie and Sverre Aarseth for useful discussions and support.

\bibliographystyle{mn2e}
\bibliography{literatur}

\bsp

\label{lastpage}

\end{document}